\documentclass[showpacs,prl,twocolumn,fleqn]{revtex4}
\usepackage{amssymb}
\usepackage{amsmath}

\input{tcilatex}

\begin{document}

\title{ Spin Hall effect and Berry phase in two dimensional electron gas }
\author{Shun-Qing Shen}
\affiliation{Department of Physics, The University of Hong Kong, Pukfulam Road, Hong
Kong, China}
\date{October 10, 2003}

\begin{abstract}
The spin Hall effect is investigated in a high mobility two dimensional
electron system with the spin-orbital coupling of both the Rashba and the
Dresselhaus types. A spin current perpendicular to the electric field is
generated by either the Rashba or the Dresselhaus coupling. The spin Hall
conductance is independent of the stength of the coupling, but its sign is
determined by the relative ratio of the two couplings. The direction of spin
current is controllable by tuning the magnitude of the\ surface electric
field perpendicular to the two dimensional plane via adjusting the Rashba
coupling. It is observed that the spin Hall conductance has a close relation
to the Berry phase of conduction electrons.
\end{abstract}

\maketitle

\textit{Introduction.}-- Spintronics in semiconductor physics has become an
emerging field of condensed matter because of its potential application in
information industry, and also because of many essential questions on
fundamental physics of electron spin.\cite{Prinz98science,Wolf01science} It
is tempting to use spin rather than charge of electron for information
precessing and storage. Recently a surprising effect was predicted
theoretically that an electric field can generate a dissipationless quantum
spin current perpendicular to the charge current at room temperatures in
conventional hole-doped semiconductors such as Si, Ge, and GaAs.\cite%
{Murakami03science} The effect was found to be intrinsic in electron systems
with substantial spin-orbit coupling, and the spin Hall conductance has a
universal value.\cite{Sinova03xxx} Based on this effect it is possible to
realize spin injection in paramagnetic semiconductors rather than from the
spin polarized carriers in ferromagnetic metals,\cite{Ohno99nature} which is
thought as a key step to realize practical spintronic devices. The spin Hall
effect due to magnetic impurities\ was discussed extensively.\cite%
{Dyakonov71,Hirsch99,Zhang00prl,Hu03prb} When a charge current circulates in
a paramagnetic metal a transverse spin imbalance will be generated, which
gives rise to a spin Hall voltage or a spin current.

In two dimensional (2D) semiconductor heterostructures the spin-orbit
interaction can be described in terms of two dominant contributions to the
model Hamiltonian, which are directly related to symmetry of the low
dimensional geometry. One type is Rashba coupling stemming from the
structure inversion asymmetry of confining potential,\cite{Rashba60} 
\begin{equation}
H_{R}=-\lambda \sigma \cdot (z\times \mathbf{k})=-\lambda (k_{x}\sigma
_{y}-k_{y}\sigma _{x})
\end{equation}%
where $\sigma _{a}$ ($\alpha =x,y,z$) are the Pauli matrices and $k_{x}$ and 
$k_{y}$ are two components of the wave vector. The strength of the coupling $%
\lambda $ can be modified by a gate field up to 50$\%$,\cite{Nitta97prl}
such that it can be applied to control spin transport, such as in spin field
transistor.\cite{Datta90apl,Shen03} Another type is Dresselhaus coupling
from the bulk inversion asymmetry,\cite{Dresselhaus54} 
\begin{equation}
H_{D}=-\beta \left( k_{x}\sigma _{x}-k_{y}\sigma _{y}\right) .
\end{equation}%
In some materials such as GaAs the two types of spin-orbit coupling are
usually of the same order of magnitudes. The interplay of them has been
investigated theoretically with respect to several phenomena, such as the
nonballistic spin field effect transistor\cite{Schliemann03prl} and
electron-spin manipulation\cite{Rashba03prl}.

In this letter we investigate the spin Hall effect in a 2D electron system
with spin-orbit coupling of both Rashba and Dresselhaus types. Consider the
Hamiltonian,%
\begin{equation}
H_{0}=\frac{\hbar ^{2}}{2m}\mathbf{k}^{2}+H_{R}+H_{D}.
\end{equation}%
We find the Berry phase in the eigenstates to be $0$ for $\left\vert \lambda
\right\vert =\left\vert \beta \right\vert $, $+\pi $ for $\left\vert \lambda
\right\vert >\left\vert \beta \right\vert ,$ and $-\pi $ for $\left\vert
\lambda \right\vert <\left\vert \beta \right\vert .$ When the system is
subjected to an electric field in the 2D plane the spin-orbit coupling leads
to a dissipationless spin current perpendicular the electric field, and
polarized in the direction perpendicular to the 2D plane when $\left\vert
\lambda \right\vert \neq \left\vert \beta \right\vert .$ The spin Hall
conductivity has a universal value except that its sign is determined by the
relative ratio of two couplings or the sign of the Berry phase. As Rashba
coupling is tunable by a gate field perpendicular to the 2D plane it is
possible to control the direction of spin current by adjusting the magnitude
rather than direction of the gate field when the system is near $\left\vert
\lambda \right\vert =\left\vert \beta \right\vert .$

\textit{Eigenstates of H}$_{0}$\textit{\ and the Berry phase.}-- The system
with two types of coupling has been investigated by several authors. The
Hamiltonian can be diagonalized exactly. The eigenstates are 
\begin{subequations}
\label{spin-state}
\begin{eqnarray}
\left\vert k_{+},+,\theta \right\rangle &=&\frac{1}{\sqrt{2}}\left( 
\begin{array}{c}
e^{-i\theta } \\ 
+i%
\end{array}%
\right) \otimes \left\vert k_{+}\right\rangle , \\
\left\vert k_{-},-,\theta \right\rangle &=&\frac{1}{\sqrt{2}}\left( 
\begin{array}{c}
e^{-i\theta } \\ 
-i%
\end{array}%
\right) \otimes \left\vert k_{-}\right\rangle .
\end{eqnarray}%
where $\left\vert k\right\rangle $ is the eigenket of k, and $\theta $ is
given by 
\end{subequations}
\begin{equation}
\tan \theta =\frac{\lambda k_{y}-\beta k_{x}}{\lambda k_{x}-\beta k_{y}}=%
\frac{\lambda \sin \varphi -\beta \cos \varphi }{\lambda \cos \varphi -\beta
\sin \varphi }
\end{equation}%
with $k_{x}=k\cos \varphi $ and $k_{y}=k\sin \varphi .$ We choose $k_{\pm }$
such that two eigenstates are degenerated%
\begin{equation}
E(k)=\frac{\hbar ^{2}}{2m}k_{+}^{2}-k_{+}\Delta \omega (\varphi )=\frac{%
\hbar ^{2}}{2m}k_{-}^{2}+k_{-}\Delta \omega (\varphi )
\end{equation}%
with $\Delta \omega \left( \varphi \right) =\sqrt{\lambda ^{2}+\beta
^{2}-2\lambda \beta \sin 2\varphi }$. In general the two bands do not cross
over except at $k=0$. In the case of $\lambda =\pm \beta ,$ $\theta =-\pi /4$
if $\lambda =\beta $ and $\pi /4$ if $\lambda =-\beta .$ The spin states in
Eq.(\ref{spin-state}) is independent of $k$ and $\varphi .$ Furthermore $%
k\Delta \omega =0$ when $k_{x}=k_{y}$ or $\varphi =\pi /4.$ It is noted that
there exists an additional conserved quantity, $\left( \sigma _{x}\pm \sigma
_{y}\right) /\sqrt{2}$.\cite{Schliemann03prl} The difference of two $k_{\pm
} $ is given by%
\begin{equation}
k_{+}\left( \varphi \right) -k_{-}(\varphi )=\frac{2m}{\hbar ^{2}}\Delta
\omega (\varphi )  \label{k-difference}
\end{equation}%
which is independent of $k.$ The two energy bands are plotted in Fig. 1. In
these states the Berry phase,\cite{Berry84} which is acquired by a state
upon being transported around a loop in the k space, can be evaluated
exactly,

\begin{equation}
\gamma _{\pm }=\doint dl\cdot \left\langle k_{\pm },\pm ,\theta \right\vert i%
\frac{\partial }{\partial \mathbf{k}}\left\vert k_{\pm },\pm ,\theta
\right\rangle =\frac{\lambda ^{2}-\beta ^{2}}{\left\vert \lambda ^{2}-\beta
^{2}\right\vert }\pi
\end{equation}%
if $\left\vert \lambda \right\vert \neq \left\vert \beta \right\vert $, and
is equal to zero if $\left\vert \lambda \right\vert =\left\vert \beta
\right\vert .$ The disappearance of the Berry phase at $\left\vert \lambda
\right\vert =\left\vert \beta \right\vert $ may be relevant to the crossover
of two energy bands. We notice that if we make a replacement, $\lambda
\rightarrow \beta $ and $\beta \rightarrow \lambda $ the angle $\theta
\rightarrow \pi /2-\theta .$ This property gives different Berry phases for $%
\left\vert \lambda \right\vert >\left\vert \beta \right\vert $ and $%
\left\vert \lambda \right\vert <\left\vert \beta \right\vert .$ According to
the theory of Sundaram and Niu,\cite{Sundaram99} the Berry phase is closely
related to the electron transport in semiconductors. It is found that the
Berry phase and magnetic monopoles in momentum space generate anomalous Hall
effect in some ferromagnetic materials.\cite{Fang03}

\FRAME{ftbpFU}{2.8556in}{3.5924in}{0pt}{\Qcb{Two energy bands: the outer
shell is for $k_{+}$-band, and the inner shell is for $k_{-}$-band. The
particle mass m is 0.067 in unit of the bare electron mass m$_{0}.$ $k$ is
in the unit of nm$^{-1}.$ (a) $\protect\lambda =5.0\times 10^{-3}$eVnm and $%
\protect\beta =7.5\times 10^{-3}$eVnm or $\protect\lambda =7.5\times 10^{-3}$%
eVnm and $\protect\beta =5.0\times 10^{-3}$eVnm. The two cases give the same
shape of energy spectra, but the corresponding spin states are different by $%
\protect\theta \rightarrow \protect\pi /2-\protect\theta .$ The two bands
coincides at the original point. (b). $\protect\lambda =\protect\beta =5.0$
meVnm The two bands cross over at $k_{x}=k_{y}$ or $\protect\varphi =\protect%
\pi /4,5\protect\pi /4.$}}{}{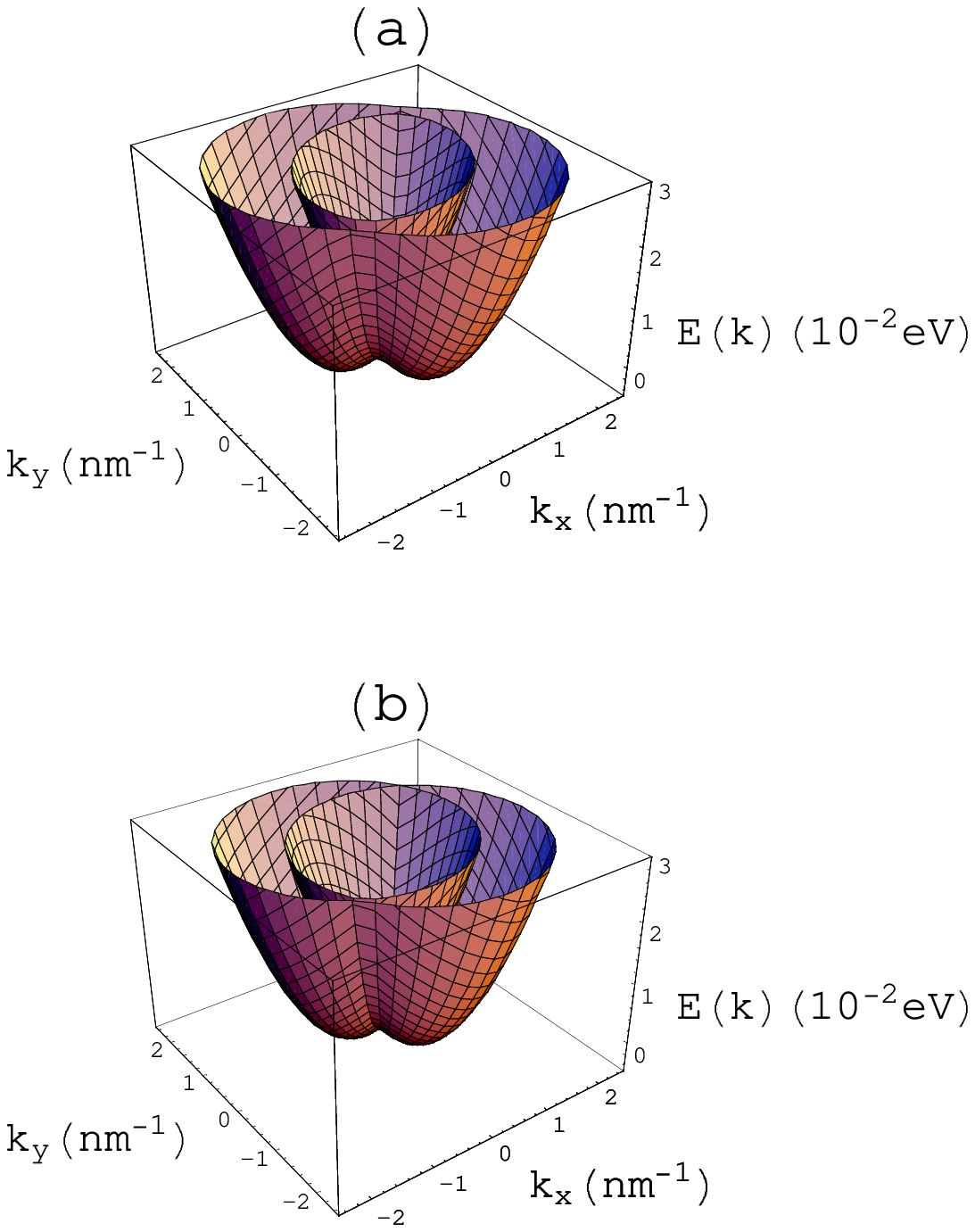}{\special{language "Scientific
Word";type "GRAPHIC";maintain-aspect-ratio TRUE;display "USEDEF";valid_file
"F";width 2.8556in;height 3.5924in;depth 0pt;original-width
4.3439in;original-height 5.4864in;cropleft "0";croptop "1";cropright
"1";cropbottom "0";filename 'xxx.eps';file-properties "XNPEU";}}

\textit{Quantum dynamics in an electric field.}-- We shall consider the
effect of uniform electric field $E$. Our whole Hamiltonian is thus given by 
$H=H_{0}+eEx$, assuming that the electric field is along the x-direction. We
start with the Heissenberg equation of motion, $i\hbar \frac{d}{dt}O=[O,H]$.
The wave vector is determined by $k_{x}(t)=k_{x0}-\frac{eE}{\hbar }t$ and $%
k_{y}(t)=k_{y0}$ where $k_{y0}$ and $k_{y0}$ are the initial values at $t=0$%
. The set of equations for spin operators are given by 
\begin{subequations}
\begin{eqnarray}
\frac{\partial }{\partial t}\sigma _{x} &=&-\frac{2}{\hbar }(\lambda
k_{x0}-\beta k_{y0})\sigma _{z}+\frac{2\lambda eE}{\hbar ^{2}}t\sigma _{z} \\
\frac{\partial }{\partial t}\sigma _{y} &=&-\frac{2}{\hbar }(\lambda
k_{y0}-\beta k_{x0})\sigma _{z}-\frac{2\beta eE}{\hbar ^{2}}t\sigma _{z} \\
\frac{\partial }{\partial t}\sigma _{z} &=&+\frac{2}{\hbar }(\lambda
k_{x0}-\beta k_{y0})\sigma _{x}+\frac{2}{\hbar }(\lambda k_{y0}-\beta
k_{x0})\sigma _{y}  \notag \\
&&-\frac{2\lambda eE}{\hbar ^{2}}t\sigma _{x}+\frac{2\beta eE}{\hbar ^{2}}%
t\sigma _{y}
\end{eqnarray}%
When $E=0$, the equations can be solved exactly 
\end{subequations}
\begin{subequations}
\begin{eqnarray}
\sigma _{x}^{0}(t) &=&\left( \cos ^{2}\theta \sigma _{x}(0)+\sin \theta \cos
\theta \sigma _{y}(0)\right) \cos \omega t \\
&&-\cos \theta \sigma _{z}(0)\sin \omega t+\sin ^{2}\theta \sigma
_{x}(0)-\sin \theta \cos \theta \sigma _{y}(0);  \notag \\
\sigma _{y}^{0}(t) &=&\left( \sin \theta \cos \theta \sigma _{x}(0)+\sin
^{2}\theta \sigma _{y}(0)\right) \cos \omega t \\
&&-\sin \theta \sigma _{z}(0)\sin \omega t-\sin \theta \cos \theta \sigma
_{x}(0)+\cos ^{2}\theta \cos \theta \sigma _{y}(0);  \notag \\
\sigma _{z}^{0}(t) &=&\sigma _{z}(0)\cos \omega t+(\cos \theta \sigma
_{x}(0)+\sin \theta \sigma _{y}(0))\sin \omega t.
\end{eqnarray}%
The characteristic frequency $\omega =\frac{2}{\hbar }k\Delta \omega
(\varphi ).$ Until now we have no general solution for the problem in the
presence of an electric field. In this letter we are concerned with the
linear response of the transport properties to the field. For our purpose we
only need an asymptotic solution for a weak field $E$ and a shot instant $t$%
. We expand the spin operators in terms of $E$: $\sigma _{a}(t)=\sigma
_{\alpha }^{0}(t)+\Delta \sigma _{\alpha }^{1}(t)+\Delta \sigma _{\alpha
}^{2}(t)+\cdots .$ For a short instant $t$, we have an asymptotic solution
for the linear correction to $\sigma _{\alpha }^{0}(t)$ due to the electric
field, 
\end{subequations}
\begin{subequations}
\label{correction}
\begin{eqnarray}
\Delta \sigma _{x}^{1} &\approx &+\frac{eE}{2k^{2}}\frac{\lambda \sigma
_{z}^{0}(t)}{\left[ \Delta \omega \left( \varphi \right) \right] ^{2}}; \\
\Delta \sigma _{y}^{1} &\approx &-\frac{eE}{2k^{2}}\frac{\beta \sigma
_{z}^{0}(t)}{\left[ \Delta \omega \left( \varphi \right) \right] ^{2}}; \\
\Delta \sigma _{z}^{1} &\approx &-\frac{eE}{2k^{2}}\frac{\lambda \sigma
_{x}^{0}(t)-\beta \sigma _{y}^{0}(t)}{\left[ \Delta \omega \left( \varphi
\right) \right] ^{2}}.
\end{eqnarray}%
We see that the effective field caused by the spin-orbit coupling has a
non-zero correction to the spin operators even when $t\rightarrow 0.$

\textit{Spin current and universal spin Hall conductance.}-- Of cause the
electric field will induce a charge current in an electron gas. The problem
has already been studied extensively\cite{Schliemann03xxx}. Spin-orbit
coupling leads to a non-zero Hall resistance proportional to $\lambda $ and $%
\beta ,$ $\rho _{xy}\propto sign(\lambda )\left\vert \lambda \beta
\right\vert .$ In this letter we focus on the spin Hall effect. We notice
that the electric field causes a linear correction to spin operators in Eqs.(%
\ref{correction}) when $t\rightarrow 0.$ This correction will generate the
spin Hall effect as discussed by Murakami \textit{et al.}\cite%
{Murakami03science} and Sinova \textit{et al.}\cite{Sinova03xxx}. Spin
current, polarized to perpendicular the 2D plane and flowing perpendicular
to the electric field,\cite{Note1} is defined as 
\end{subequations}
\begin{equation}
j_{y}^{z}=\frac{\hbar }{4}\left\langle \left\{ \sigma _{z}(t),\frac{\partial 
}{\partial t}y(t)\right\} \right\rangle .
\end{equation}%
where $\frac{\partial }{\partial t}y=\frac{\hbar }{m}k_{y0}+\frac{\lambda }{%
\hbar }\sigma _{x}(t)+\frac{\beta }{\hbar }\sigma _{y}(t).$ The expectation
value is taken over the energy eigenstates of electrons. The linear term to
the electric field is given by%
\begin{eqnarray}
j_{y}^{z} &=&\sum_{k}\frac{\hbar ^{2}}{2m}\left\langle k_{+},+,\theta
\right\vert \Delta \sigma _{z}^{1}(t)k_{y0}\left\vert k_{+},+,\theta
\right\rangle n_{F}(E(k_{+})-\mu )  \notag \\
&&+\sum_{k}\frac{\hbar ^{2}}{2m}\left\langle k_{-},-,\theta \right\vert
\Delta \sigma _{z}^{1}(t)k_{y0}\left\vert k_{-},-,\theta \right\rangle
n_{F}(E(k_{+})-\mu )
\end{eqnarray}%
where $n_{F}(E(k))$ is the Dirac-Fermi distribution. At zero temperature we
have%
\begin{equation*}
j_{y}^{z}=\frac{\hbar ^{2}eE}{16\pi ^{2}m}\int_{0}^{2\pi }d\varphi \frac{%
\left( \lambda ^{2}-\beta ^{2}\right) \sin ^{2}\varphi \left[ k_{+}-k_{-}%
\right] }{(\lambda ^{2}+\beta ^{2}-2\lambda \beta \sin 2\varphi )^{3/2}}.
\end{equation*}%
By using the relation in Eq.(\ref{k-difference}) we obtain the spin Hall
conductance 
\begin{equation}
\sigma _{sH}=j_{y}^{z}/E=\frac{e}{8\pi ^{2}}\gamma _{\pm }  \label{linear}
\end{equation}%
as long as both bands are occupied.\cite{Note2} We have ignored the
time-dependent terms. It is an interesting observation that the spin Hall
conductance is closely relevant to the Berry phase of electron states in the
absence of the electric field. It deserves for further investigation on the
relation in general cases. Spin current caused by the Berry phase should not
contribute to dissipation, although a charge current will do. If $\beta =0$ $%
\sigma _{sH}=e/8\pi $ which recovers Sinova \textit{et al.}'s result\cite%
{Sinova03xxx} and followed by several authors more recently\cite%
{Schliemann03yyy,Hu03xxx}$.$ If $\lambda =0,$ Dresselhaus coupling can also
produce a spin Hall conductance $\sigma _{sH}=-e/8\pi $. Very interestingly
two spin Hall conductances differ from only a sign. Competition of two
couplings also cancel the effect at $\lambda =\pm \beta .$ It seems that the
effect caused by Dresselhaus coupling dominates if $\left\vert \beta
\right\vert >\left\vert \lambda \right\vert $ while the effect caused by
Rashba coupling dominates if $\left\vert \lambda \right\vert >\left\vert
\beta \right\vert $. Very recently Rashba\cite{Rashba03xxx} observed that
even in the absence of an external electric field there exist a tiny spin
current $J_{y}^{x}=-J_{x}^{y}\neq 0$ in the open boundary condition. The
type of spin current along spin z-direction does not exist in the absence
and presence of an electric field.

To explore the physical origin of the sign change of spin Hall conductance
near $\lambda =\beta $ we find a new symmetry in the system with the two
spin-orbital couplings. Performing an unitary transformation, 
\begin{equation}
\sigma _{x}\rightarrow \sigma _{y};\text{ }\sigma _{y}\rightarrow \sigma
_{x};\text{ }\sigma _{z}\rightarrow -\sigma _{z},
\end{equation}%
the two terms of Rashba and Dresselhaus couplings are changed to 
\begin{subequations}
\begin{eqnarray}
H_{R} &=&-\lambda (k_{x}\sigma _{y}-k_{y}\sigma _{x})\rightarrow -\lambda
(k_{x}\sigma _{x}-k_{y}\sigma _{y}); \\
H_{D} &=&-\beta (k_{x}\sigma _{x}-k_{y}\sigma _{y})\rightarrow -\beta
(k_{x}\sigma _{y}-k_{y}\sigma _{x}).
\end{eqnarray}%
The spin current along the z-direction changes a minus sign, $%
J_{y}^{z}\rightarrow -J_{y}^{z}$ as $\sigma _{z}\rightarrow -\sigma _{z}.$
Thus this symmetry indicates that the system with Rashba coupling $\lambda $
and Dresselhaus coupling $\beta $ and the system with Rashba coupling $\beta 
$ and Dresselhaus coupling $\lambda $, have the same spin Hall conductance
(along spin z direction), but with opposite signs. At the symmetric point $%
\lambda =\beta $, we conclude that $J_{y}^{z}=0$, i.e., the spin Hall effect
is suppressed completely. This general conclusion is based on symmetry
analysis of the system and is mathematically rigorous. It is in agreement
with the result of the linear response calculation in Eq. (\ref{linear}).

Now we come to address the effects of a finite electron quasi-particle
lifetime and scattering mechanism, which was discussed in a system with
Rashba coupling by several authors. Schliemann and Loss\cite{Schliemann03yyy}
discussed the correction of spin conductance due to the effect of finite
lifetime $\tau $ for $\beta =0$ using the Kubo formula in the zero frequency
limit and concluded that the universal value for spin conductance is valid
only when the \textquotedblleft Rashba energy\textquotedblright\ $%
\varepsilon _{R}=m\lambda ^{2}/\hbar ^{2}$ should be comparable with the
energy scale $\hbar /\tau .$ Fortunately in some semiconductors such as GaAs
quantum wells $\hbar /\tau <\varepsilon _{R}$ and it is safely to neglect
these effects. On the other hand, more recently Burkov and MacDonald found
that the spin Hall conductance in a system with Rashba coupling remains the
universal value $e/8\pi $ in the case of strong disordering as in the case
of weak disordering.\cite{Burkov03xxx} Thus the scattering effect depends on
the scattering mechanisms.

\textit{Controllable Rashba coupling and revertible spin current.}-- Rashba
coupling and Dresselhaus coupling are related to the symmetry of quantum
wells. In GaAs quantum wells both terms are usually of the same order of
magnitude while the narrow gap compounds like in InAs Rashba coupling
dominates.\cite{Jusserand95} However there is no difficulty to achieve the
situation of $\lambda =\beta $. In the case of $\lambda =\pm \beta $ the
Elliot-Yafet spin flip mechanism is suppressed completely since the spin
states in $\left\vert k_{\pm },\pm ,\theta =\pm \pi /4\right\rangle $ is
independent of $\varphi $. The Berry phase of the states $\left\vert k_{\pm
},\pm ,\theta \right\rangle $ becomes zero and correspondingly there is no
spin Hall effect. Thus the electric field cannot generate a dissipationless
spin current. If $\left\vert \lambda \right\vert >\left\vert \beta
\right\vert $ $\sigma _{sH}=e/8\pi $ while $\sigma _{sH}=-e/8\pi $ if $%
\left\vert \lambda \right\vert <\left\vert \beta \right\vert .$ The spin
current changes a sign (or direction). As Rashba coupling is adjustable by a
gate field perpendicular to the electron gas plane\cite{Nitta97prl} we may
change $\gamma $ around $\beta ,$ $\gamma =\beta +\delta \Delta E_{\bot },$
by adjusting the perpendicular electric field $\Delta E_{\bot }$. Thus by
using the property of sign change in the Berry phase we can realize to
revert the spin current by adjusting the magnitude rather than the direction
of the gate field.

\textit{Conclusion.}-- The spin Hall conductance caused by Dresselhaus
coupling has the same universal value, but an opposite sign to that caused
by Rashba coupling. It is an interesting observation that the spin Hall
conductance has a close relation to the Berry phase of single electron
states. The spin current in this effect can be revertible when the
amplitudes of two couplings are comparable by adjusting the gate voltage via
Rashba coupling. It is anticipated that this effect will be applicable to
the spintronic devices in the future.

The author thanks S. Murakami, Q. Niu, J. Sinova, X. C. Xie, and F. C. Zhang
for discussions. This work was supported by a grant from the Research Grant
Council of Hong Kong (Project No.: HKU 7088/01P). After this paper was
submitted for publication the author was aware that Sinova \textit{et al.}
arrived at the same spin Hall conductances by means of the linear response
theory independently.\cite{Sinitsyn03xxx}

\end{subequations}

\end{document}